# Impact of Metro Cell Antenna Pattern and Downtilt in Heterogeneous Networks

Xiao Li, Southeast University, and The State Key Laboratory of Integrated Services Networks; Robert W. Heath, Jr., The University of Texas at Austin; Kevin Linehan and Ray Butler, CommScope

*This article discusses the positive impact metro cell antennas with narrow vertical beamwidth and electrical downtilt can have on heterogeneous cellular networks. Using a model of random cell placement based on Poisson distribution, along with an innovative 3D building model that quantifies blockage due to shadowing, it is demonstrated that network spectral efficiency and average user throughput both increase as vertical beamwidth is decreased and downtilt is applied to metro cell transmission. Moreover, the network becomes more energy efficient. Importantly, these additional gains in network performance can be achieved without any cooperation or exchange of information between macro cell base stations and metro cells.*

## Introduction

It is widely acknowledged that wireless networks continue to face significant growth in user data demands as media-hungry devices and applications are adopted. The traditional approach for improving network performance is to either increase area spectral efficiency through cell splitting or add new cell sites. However, network expansion using macro cellular infrastructure is costly and usually restricted by zoning. A rising alternative is the deployment of heterogeneous networks [1], [2] in which various low-power nodes—often referred to as "small cells" (which include femto cells, pico cells, metro cells, and distributed antennas systems)—are added. Similar to the placement of macro cellular infrastructure, in which sites are distributed based often on urban sprawl and site availability, small cells can also be expected to be distributed in a somewhat random manner: being deployed where possible and immediately beneficial instead of in optimum locations. This random placement of infrastructure driven by forces outside the mobile network operator's control challenges the notion of using uniform, hexagonal base station network models for analysis. Alternatively, use of a random distribution model according to certain spatial distribution and stochastic geometry [3]-[5] can provide us with a way to obtain simple closed-form performance characterizations of traditional cellular [3] and heterogeneous networks [4], [5] without running extensive simulations. This technique can also be extended to study coordinated cellular networks [6] as well as millimeter wave cellular networks [7].

The purpose of this article is to illustrate the impact antenna downtilt and vertical beamwidth can have on heterogeneous macro/metro cellular networks. A stochastic model of the 3D environment is used to evaluate network performance. This model takes into account both horizontal and vertical antenna radiation patterns, as well as blockage due to buildings in the path of transmission. The performance impact of using electrical downtilt in traditional hexagonal 3D cellular networks was investigated in [8]-[10]. However, no such 3D model for metro cells in heterogeneous networks has existed to date. In this article, we consider a two-tier random heterogeneous cellular network and describe a simulation platform adopting horizontal sectors in macro cells and downtilted 3D antenna patterns among both macro and metro cells. This enables us to incorporate antenna parameters such as half-power beamwidth (HPBW), side-lobe suppression, horizontal sectoring, and antenna downtilt into our simulation. We then apply a 3D random building model to the simulation in order to quantify blockage due to shadowing. Based on this, we investigate the impact of metro cell antenna vertical HPBW and downtilt on the area spectral efficiency and average user rate. As demonstrated, incorporation of antennas with narrow vertical beamwidth and optimal downtilt into the metro cell layer of a heterogeneous network causes it to perform better and become more energy efficient. Importantly, this can occur without any cooperation or exchange of information between network elements.

## Network Topology

Traditionally, cellular networks are modeled by placing the macro cell base stations (BSs) on a hexagonal grid with each BS serving 3-6 cells. A heterogeneous network model was proposed in [11] based on this hexagonal grid macro cell topology. In this model, there are six metro cells per macro cell, and the metro cells are located precisely on the boundaries between neighboring macro cell BSs. Despite their extensive use, these kinds of perfect grid models are idealized relative to actual BS deployment and lack tractability [2].

Using stochastic geometry, a tractable approach was proposed in [3] to characterize the performance of traditional cellular networks. In this approach, the BS locations are chosen from a homogeneous Poisson point process (PPP), and each user is associated to the nearest BS. This approach was then extended to heterogeneous networks by modeling different tiers of BSs as independent PPPs [4], [5]. It was shown in [3] that the PPP model provides a lower bound of expected performance versus the actual BS deployment, while the grid model provides an upper bound. However, there were some important limitations in this model. First, it lacked an ability to model sectored antennas, which are common to macro cellular infrastructure; and second, everything was designed on a 2D plane. Lastly, the propagation environment relied on simple, distance-dependent path-loss models combined with small-scale fading.

To overcome the limitations of previous PPP models, we use a two-tier heterogeneous 3D cellular network model (Fig. 1) in which the first tier holds the BSs of the macro cells and the second tier holds the metro cells. In this model, the wireless access points (WAPs) in the i-th tier are spatially distributed as a PPP $\Phi_i$ of density $\lambda_i$ (the WAPs are all independent and uniformly distributed, the number of WAPs in the i-th tier is a Poisson variable with an average $\lambda_i$ on unit area) and transmit at power $P_i$. A mobile can reliably communicate with a WAP $X$ in the i-th tier only if its downlink signal-to-interference-plus-noise ratio (SINR) with respect to that WAP is greater than $\beta_i$. Both WAP tiers (macro and metro) have certain height and 3D directional antenna patterns. In this model, each macro BS serves three sectors, similar to the traditional grid model, and the horizontal direction of the main beam of each sector is random and uniformly distributed in [0,2π), although the horizontal angle between the main beams of two adjacent sectors in the same macro cell is fixed to 120°. There is no information exchange between WAPs and adjacent sectors, and no interference coordination between different WAPs. Therefore, signals from all other WAPs and sectors other than the home WAP or sector are treated as interference. This scenario is actually the worst-case network and is used to understand network performance and insight into tradeoffs in network design. Furthermore, there is no intra-cell interference, due to orthogonal multiple access within each cell. Figure 2 shows an example of the cell edge boundaries of this 3D heterogeneous network without the impact of buildings (shadow fading). The red points represent macro cell BSs and the green diamonds represent metro cells. The boundaries are defined by SINR, i.e., users associate with the WAP that offers the best SINR (with some SINR bias between different tiers).

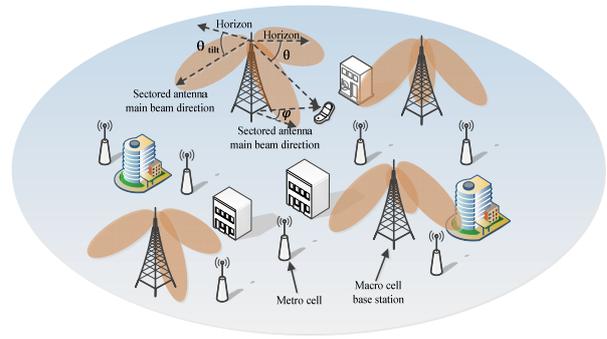

*Figure 1* 3D heterogeneous network model.

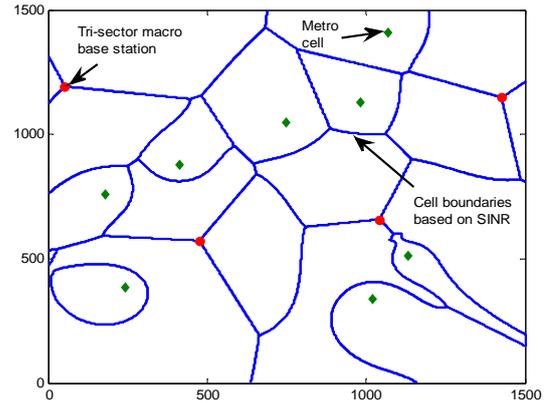

*Figure 2* Example of the cell boundaries for our 3D heterogeneous network.

A narrowband, frequency-flat fading channel and single antenna is assumed to be used by both WAPs and mobile users. In this case, the channel coefficient between WAP $X$ and the user can be written as [8]:

$$h_X = \sqrt{S \cdot G(\varphi,\theta,\theta_{\text{tilt}}) \cdot L^{-1}(d)} h_{wX}, \qquad (1)$$

where $S=\gamma^{K_x}$ corresponds to shadow fading [12], $K_x$ is the random number of buildings that cross the direct path between WAP $X$ and the user, $\gamma<1$ is the attenuation coefficient of each building, $L^{-1}(d)$ is the path-loss, $d$ is the distance between WAP $X$ and the user in kilometers, and $h_{wX}$ denotes the small-scale fading, which is assumed to be Rayleigh distributed. Here, $G(\varphi,\theta,\theta_{\text{tilt}})$ is the 3D antenna gain (we will describe it in detail later), $\varphi$ is the horizontal angle relative the main beam pointing direction (see Fig. 1), $\theta_{\text{tilt}}$ is the electrical downtilt angle of WAP's antenna, $-90° \leq \theta \leq 90°$ is the negative elevation angle relative to the horizontal plane. Note that the downtilt could be mechanical or electrical. However, in industry, electrical tilting is more efficient and preferable than mechanical tilting [13, Section 6.2]. Therefore, in this paper, we consider only electrical downtilt. Moreover, for the horizontal omni antenna we considered for the metro cells, mechanical downtilt is not applicable.

## Incorporating a 3D Building Model

In this stage, a 3D building model based on the 2D random building model in [12] is incorporated. The buildings are modeled as rectangles using the building centers on the 2D plane; the height and length of the rectangle are the height and length of the building; and the angle between the rectangle and x-axis on the 2D plane is the orientation of the building (see Fig. 3). We assume that the building centers are distributed as a homogenous PPP $\Phi_B$ with building density $\lambda_B$. The length of the building with center $X_B$ is distributed with probability density function $f_l(x)$, its orientation is uniformly distributed between $[0,2\pi)$, and its height is distributed with probability density function $f_h(x)$. Note that $f_l(x)$ and $f_h(x)$ could be any appropriate distribution function.

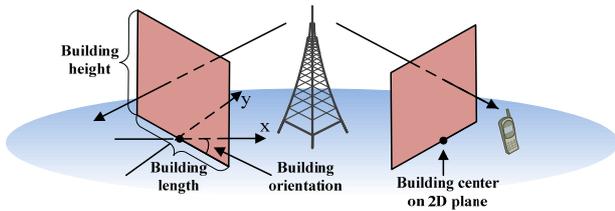

**Figure 3** *3D building model.*

## Applying 3D Antenna Patterns

3D antenna patterns are assumed at both the macro cell BSs and metro cells. According to [14] and [15], the 3D antenna gain $G(\varphi,\theta,\theta_{tilt})$ in (1) can be approximated in dBi as:

$$G(\varphi,\theta,\theta_{tilt})|_{dBi} = G_h(\varphi)|_{dBi} + G_v(\theta,\theta_{tilt})|_{dBi} + G_m|_{dBi}, \quad (2)$$

where $G_h(\varphi)|_{dBi}$ and $G_v(\theta,\theta_{tilt})|_{dBi}$ are the normalized horizontal and vertical antenna gain in dBi, respectively, and $G_m|_{dBi}$ is the maximum antenna gain in dBi.

For the macro cell BSs, we use the 3D antenna pattern for each sector defined in [15], where $G_h(\varphi)|_{dBi} = -\min[12(\varphi/B_h)^2, F_h]$ and $G_v(\theta,\theta_{tilt})|_{dBi} = \max[-12((\theta-\theta_{tilt})/B_{v1})^2, F_{v1}]$ with electrical downtilt, $B_h$ and $B_{v1}$ are the horizontal and vertical HPBW, respectively, $F_h$ is the horizontal front back ratio (FBR), $F_{v1}$ is the vertical side-lobe level (SLL).

For the metro cells, we consider the horizontal omni antenna [1], i.e., $G_h(\varphi)|_{dBi} = 0$ dBi. For the vertical pattern, we consider the dipole antennas. With electrical downtilt, the vertical pattern of the dipole antenna main lobe can be approximated by [16] $G_v(\theta,\theta_{tilt})|_{dBi} = 10\log_{10}|\cos^n(\theta-\theta_{tilt})|$, where $n$ relates to the antenna vertical HPBW. In this work, we further take into account the vertical SLL, and approximated the vertical antenna gain as $G_v(\theta,\theta_{tilt})|_{dBi} = \max[10\log_{10}|\cos^n(\theta-\theta_{tilt})|, F_{v2}]$, where $F_{v2}$ is the vertical SLL. Denote the vertical HPBW of the metro cell antenna as $B_{v2}$. From the definition we can get that $G_v(B_{v2}/2,0)|_{dBi} = 3$ dBi. Therefore, it can be obtained that n=2.75 for a single-element (1-element) half-wave dipole antenna ($B_{v2}$=78°), n=11.73 for a 2-element half-wave dipole antenna ($B_{v2}$=39°), and n=47.64 for a 4-element half-wave dipole antenna ($B_{v2}$=19.5°). Note that, by reducing the beamwidth by half, the corresponding maximum antenna gain will be 3 dBi greater. Therefore, the $G_m|_{dBi}$ of a 2-element dipole antenna is 3 dBi greater than the $G_m|_{dBi}$ of a 1-element dipole antenna; and the $G_m|_{dBi}$ of a 4-element dipole antenna is 3 dBi greater than the $G_m|_{dBi}$ of a 2-element dipole antenna.

## User Association

In a cellular network, each user has to associate with an access point: a macro cell BS or metro cell in this instance. In [3], each user associates with the nearest macro cell BS. For a multi-tier network, different tiers of access points might have different transmit power and different coverage thresholds. While the choice of access point can be made based on SINR, using instantaneous SINR places a significant burden on feedback link, especially when the number of users is large and the users are highly mobile. Furthermore, it is hard to estimate the exact instantaneous SINR of all users in a network. In this article, we assume that the mobile users are interference limited and use long-term information: average-signal-to-average-interference ratio (ASAIR) $E\{D\}/E\{I\}$ to associate users, where $D$ is the desired signal power, $I$ is the received interference power, and $E\{\}$ denotes the expectation. From (1) it can be seen that the ASAIR can be obtained by antenna gain, path loss, etc., which varies relatively slowly. Note that both instantaneous and averaged channel state information (CSI) is applicable in user association, and many previous work associate users using instantaneous CSI, in our work we choose the averaged CSI. The user association algorithm is as follows:

1) Find the macro cell BS $X_1 \in \Phi_1$ containing the sector (denoted as j-th sector) that provides the highest average signal power among the sectors in tier 1, and calculate the corresponding ASAIR $\bar{\rho}(X_1)$.
2) Find the metro cell $X_2 \in \Phi_2$ that provides the highest average signal power among the metro cells, and calculate the corresponding ASAIR $\bar{\rho}(X_2)$.
3) If $\bar{\rho}(X_2) \geq \bar{\rho}(X_1) - \beta$, associate the user with the metro cell $X_2 \in \Phi_2$; otherwise, associate the user with the j-th sector of the macro cell BS $X_1 \in \Phi_1$. Here, $\beta = \beta_1 - \beta_2$ is the signal-to-interference ratio (SIR) threshold bias between different tiers of WAP.

Step 1 and Step 2 are based on the fact that the WAP providing the highest ASAIR in the i-th tier is the one providing the highest average signal power in the corresponding tier. Step 3 is based on the fact that the macro tier usually has a higher SIR threshold than the metro tier, i.e., $\beta$ dB higher, since macro cell BSs usually serve many more users than metro cells. We use such an algorithm that associates users with the macro tier only if $\bar{\rho}(X_1)$ is $\beta$ dB higher than $\bar{\rho}(X_2)$.

## Results

This section presents simulation results for the area spectral efficiency and average user rate performance of the downlink 3D heterogeneous cellular network described above with different metro cell antenna patterns and downtilts. In it, we compare the performance of four metro cell antenna patterns: a 1-element half-wave dipole, a 2-element half-wave dipole, a 4-element half-wave dipole, and a horizontal "quasi" omnidirectional antenna from CommScope. The quasi-omni antenna shares several characteristics with its larger macro cell brethren:

- Multiple elements are used to narrow the vertical beamwidth, resulting in a vertical HPBW of 14°-19°, depending on frequency.
- It contains integral electrical downtilt with upper side-lobe suppression.
- It is comprised of three combined 65° azimuth beamwidth panels phased appropriately to act in a quasi-omnidirectional mode. It can be reconfigured for sectorization in the future, as additional capacity is needed, with the addition of radios.
- It is dual-polarized.

In our work, the horizontal quasi-omni antenna we have chosen has a 14° vertical HPBW and 16 dB SLL. We model its vertical pattern as $G_v(\theta,\theta_{tilt})=\max[-12((\theta-\theta_{tilt})/14)^2, -16]$. Its horizontal pattern is shown in Fig. 4. Vertical patterns of these metro antennas are shown in Fig. 5, and the main parameters are shown in Table 1. Without loss of generality, we conduct a simulation on a typical mobile user located at the origin [3], [4].

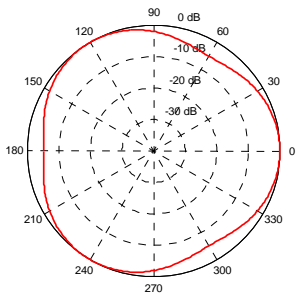

*Figure 4* Horizontal pattern of quasi-omni antenna.

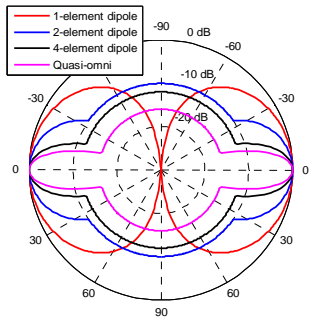

*Figure 5* Vertical patterns of different metro cell antennas.

|  | HPBW | Vertical SLL | Maximum gain |
|---|---|---|---|
| 1-element dipole | 78° | NA | 2.15 dBi |
| 2-element dipole | 39° | -10 dB | 5.15 dBi |
| 4-element dipole | 19.5° | -12 dB | 8.15 dBi |
| Horizontal quasi-omni | 14° | -16 dB | 10.2 dBi |

*Table 1* Main parameters of different metro cell antennas.

We assume a full buffer traffic model such that the user density is sufficiently large so there is at least one user scheduled per cell/sector. Therefore, the area spectral efficiency of the 3D heterogeneous network can be written as $R=3\lambda_1 R_1+ \lambda_2 R_2$, where $\lambda_1$ is the macro cell BS density, $\lambda_2$ is the metro cell density, $R_1=E\{\log_2(1+ \rho(X))|X\in\Phi_1\}$ is the average spectral efficiency of each macro cell sector, the coefficient "3" is used because the density of the macro sector is three times the density of the macro cell BS, $R_2=E\{\log_2(1+\rho(x))|x\in\Phi_2\}$ is the average spectral efficiency of each metro cell, $\rho(x)$ is the SIR when a user associates with the WAP $X$. The average rate achievable by a randomly located mobile user can be written as $R=R_1\Pr(X\in\Phi_1)+R_2\Pr(X\in\Phi_2)$. Note that for more realistic traffic model the performance would be better, since the interference would be weaker. The main simulation parameters are summarized in Table 2.

| Simulation parameter | Value |
|---|---|
| Macro Cell Tx Power $P_1$ | 46 dBm |
| Macro Cell Base Station Density $\lambda_1$ | 2.05/km$^2$ (equal to the density of hexagonal grid model with 750m inter-site distance [ISD]) |
| Macro Cell Antenna Height | 30 m |
| Horizontal HPBW of Macro Cell Antenna | 65° |
| FBR of Macro Cell Antenna | 25 dB |
| Downtilt of Macro Cell Antenna | 10° |
| Vertical HPBW of Macro Cell Antenna | 7° |
| SLL of Macro Cell Antenna | -18 dB |
| Maximum Gain of Macro Cell Antenna | 18 dBi |
| Macro Cell Pathloss Model | 3GPP TR 36.814 model 1: $L^{dB}(d)=128.1+37.6\log_{10}(d)$ |
| Metro Cell Tx Power $P_2$ | 33 dBm |
| Metro Cell Density $\lambda_2$ | 15$\lambda_1$ (5 metro nodes per macro sector) |
| Metro Cell Antenna Height | 5 m |
| Metro Cell Pathloss Model | 3GPP TR 36.814 model 1: $L^{dB}(d)=140.7+36.7\log_{10}(d)$ |
| Downtilt of Metro Cell Antenna | 0°, 8° and 16° |
| SIR Bias $\beta$ | 6 dB |
| Building Attenuation Coefficient $\gamma$ | -40 dB |
| Building Density $\lambda_B$ | 15$\lambda_1$ |
| Building Height | Uniformly distributed between 10m and 20m |

| Building Length | Uniformly distributed between 20m and 30m |
|---|---|
| Carrier Frequency | 2 GHz |

*Table 2 Main simulation parameters.*

## Same Metro Cell Transmit Power Using Different Antenna Patterns and Gains

The area spectral efficiency and average user rate performance of the downlink 3D heterogeneous cellular network versus metro cell antenna downtilt are shown in Figure 6 and Figure 7 for different metro cell antenna patterns under the same metro cell transmit power. In these figures, the metro cell transmit power $P_2$ is 33 dBm. Note that the single-element dipole antenna can not be tilted down.

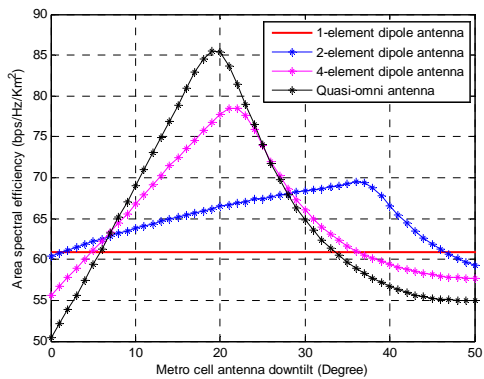

*Figure 6 Area spectral efficiency performance of the 3D heterogeneous network for different metro cell antenna patterns and downtilts under the same transmit power.*

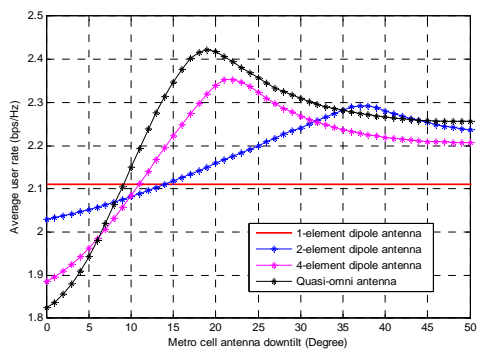

*Figure 7 Average user rate performance of the 3D heterogeneous network for different metro cell antenna patterns and downtilts under the same transmit power.*

Table 3 and Table 4 present the radius of the metro cells with a 4-element dipole antenna and quasi-omni antenna under different downtilts; the metro cell transmit power $P_2$ is also 33 dBm. The radius is measured at the outer -3 dB point of the main beam on the ground. The percentage of users served by metro cells is also given in these tables.

| Downtilt | Radius of metro cell | Percentage of users served by metro cells |
|---|---|---|
| 10° | 1145.9 m | 34.1% |
| 20° | 27.7 m | 21.7% |
| 30° | 13.6 m | 14.9% |
| 40° | 8.6 m | 14.4% |

*Table 3 Radius of metro cells and percentage of users served by metro cells for the metro cells with 4-element dipole antennas under different downtilts.*

| Downtilt | Radius of metro cell | Percentage of users served by metro cells |
|---|---|---|
| 10° | 95.4 m | 44.7% |
| 20° | 21.7 m | 22.1% |
| 30° | 11.8 m | 18% |
| 40° | 7.7 m | 17.6% |

*Table 4 Radius of metro cells and percentage of users served by metro cells for the metro cells with quasi-omni antennas under different downtilts.*

We can observe the following from Figure 6, Figure 7, Table 3, and Table 4.

- For the same metro cell antenna pattern, both the area spectral efficiency and average user rate of the 3D heterogeneous network first increase and then decrease as the antennas are tilted down. There is an optimal downtilt for each metro cell antenna pattern. This is because the interference from one metro cell to the other metro or macro cells is decreased as the metro cell antennas are tilted down. However, as the metro cell antennas are further tilted down, their covered region gets smaller and smaller, which can be observed from Table 3 and Table 4. Some users that were previously associated with metro cells may not be able to access them, and, at the same time, also suffer from poor SIR at the macro tier. Therefore, the performance degrades as the metro cell antenna downtilt keeps increasing. It can be seen from these figures that there is a floor for each curve. This represents the case that the metro cell antennas are tilted down too much, so the users associated with the metro cells can communicate only through the side lobe of the antenna. Because of this, network engineers should avoid excessive downtilt for the sake of network throughput.
- When there is no downtilt at the metro cells, both the area spectral efficiency and average user rate decrease as the antenna beamwidth decreases. This is because the gain of the antenna with smaller beamwidth is higher than the one with larger beamwidth. Therefore, the interference from the metro cell with smaller beamwidth is stronger than the one with larger beamwidth.
- When the metro cell antennas are tilted down, the trends are different from the case when there is no

downtilt; both maximum achievable area spectral efficiency and average user rate increase as the metro cell antenna beamwidth decrease. This is because the metro cells with smaller beamwidth experience higher desired signal power while causing less interference to other cells.
- The narrower the metro cell antenna vertical beamwidth, the smaller the corresponding optimal downtilt. This is because, compared to the metro cells with wider vertical antenna beamwidth, the interference from the metro cells with narrower vertical antenna beamwidth decreases faster as the downtilt increases.
- The percentage of users served by metro cells, or equivalently the covered region of metro cells, decreases as metro cell antennas are tilted down. Therefore, the network designer should trade off between the network throughput and the metro cell load. Excessive downtilt should be avoided for the sake of the covered region of metro cells.
- The heterogeneous network could achieve almost as much as 40 percent gain in area spectral efficiency and 14 percent gain in average user rate just through changing the metro cell antenna from the 1-element dipole antenna to the downtilted horizontal quasi-omni antenna, all without any cooperation or exchange of information between WAPs. For a 10° electrical downtilt, the quasi-omni antenna has a smaller cell radius, yet captures 10 percent more users than the 4-element dipole antenna and has an ASE advantage of 7 percent. At 20° of electrical downtilt the quasi-omni antenna captures roughly the same percentage of users as the 4-element dipole antenna, but has an ASE advantage of 9.3 percent.

### Same EIRP Using Different Antenna Patterns

In this section, we make the assumption that metro cell antennas with different patterns have the same EIRP, i.e., if the maximum gain increases by 3 dBi, the transmit power $P_2$ should be reduced by 3 dBm. Table 5 illustrates the maximum antenna gain and transmit power for each metro cell antenna pattern. The area spectral efficiency and average user rate performance of the downlink 3D heterogeneous cellular network are shown in Figure 8 and Figure 9 for different metro cell antenna patterns and downtilts under the same EIRP.

|  | Maximum gain | Transmit power |
|---|---|---|
| 1-element dipole | 2.15 dBi | 33 dBm |
| 2-element dipole | 5.15 dBi | 30 dBm |
| 4-element dipole | 8.15 dBi | 27 dBm |
| Horizontal quasi-omni | 10.2 dBi | 24.95 dBm |

**Table 5** *Maximum gain and transmit power for different metro cell antenna patterns.*

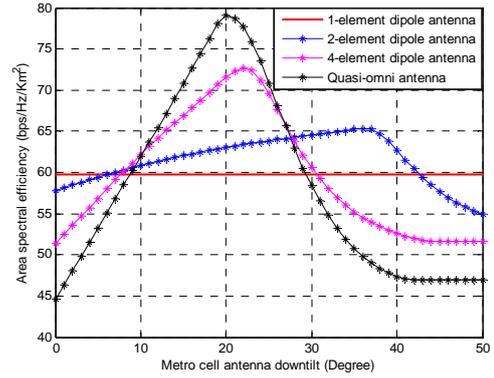

**Figure 8** *Area spectral efficiency performance of the 3D heterogeneous network with different metro cell antenna patterns and downtilts under the same EIRP.*

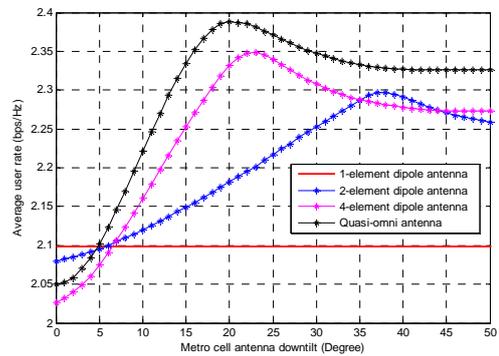

**Figure 9** *Average user rate performance of the 3D heterogeneous networks with different metro cell antenna patterns and downtilts under the same EIRP.*

We observe the following from Figure 8 and Figure 9:
- For the same metro cell antenna pattern, both the area spectral efficiency and average user rate of the 3D heterogeneous network first increase and then decrease as the antennas are tilted down. There is an optimal downtilt and performance floor for each metro cell antenna pattern. This trend was also demonstrated in the previous section.
- When there is no downtilt at the metro cell antenna, the area spectral efficiency decreases as the antenna beamwidth decreases. In this case, although the interference from the metro cell with a smaller beamwidth does not increase due to the same EIRP assumption, the desired signal power received by the metro cell user decreases, resulting in the decrease in area spectral efficiency.
- When the metro cell antennas are tilted down, both maximum achievable area spectral efficiency and average user rate increase as metro cell antenna beamwidth decreases, although the metro cell with a smaller beamwidth has less transmit power. Therefore, when metro cell antennas are tilted down to their optimal position, the heterogeneous network with a smaller metro cell antenna vertical beamwidth is more energy efficient, i.e., has higher capacity with less metro cell power.
- From Figure 8, in which the EIRP is assumed to be equal at all antennas, it can be seen that maximum area spectral efficiency can be achieved using antennas with more elements (causing narrower vertical beamwidth) and upper side-lobe suppression as illustrated in Figure 5. It is therefore demonstrated that increased gain and narrower vertical beamwidth contribute to the increased area spectral efficiency, with EIRP at the antennas being equal.
- The heterogeneous network can still achieve almost as much as 31 percent gain in area spectral efficiency and 13 percent gain in average user rate with 8.05 dBm lower metro cell transmit power just by changing the metro cell antenna from the 1-element dipole antenna to the downtilted horizontal quasi-omni antenna. The heterogeneous network with a smaller metro cell antenna vertical beamwidth is more energy efficient.

## Conclusions

In this article we introduced a 3D random heterogeneous cellular network model and investigated the impact metro cell antennas with electrical downtilt can have on heterogeneous networks. During modeling, we incorporated downtilted 3D antenna patterns in both the macro and metro cellular networks, and a novel 3D building model was adopted to quantify blockage due to shadowing. We concluded that the area spectral efficiency and average user rate of our 3D random heterogeneous network increase as the metro cell antenna vertical beamwidth decreases and the corresponding optimal downtilt is applied. This demonstrated that heterogeneous networks can achieve considerable performance gains while becoming more energy efficient through the adoption of antennas that offer superior vertical beamwidth control alone, without inter-site cooperation.

## Acknowledgements

This work is supported by A Foundation for the Author of National Excellent Doctoral Dissertation of PR China under Grant 201446, National Natural Science Foundation of China under Grant 61222102, the Excellent Young Teachers Program of Southeast University under Grant 2242015R30006, the Opening Project of The State Key Laboratory of Integrated Services Networks under Grant ISN13-03, the National Science Foundation under Grant NSF-CCF-1218338, and by CommScope.